\documentclass[prb,aps,showpacs,twocolumn,superscriptaddress,10pt,amsmath,amssymb]{revtex4-1}

\usepackage{graphicx}
\usepackage{dcolumn}
\usepackage{bm}
\usepackage{amsmath}
\usepackage{epsfig}
\usepackage{rotating}
\usepackage{enumerate}
\usepackage{xcolor}

\begin{document}

\title{Electric control of the heat flux through
       electrophononic effects}

\author{Juan Antonio Seijas-Bellido}
\affiliation{Institut de Ci\`encia de Materials de Barcelona (ICMAB--CSIC)
             Campus de Bellaterra, 08193 Bellaterra, Barcelona, Spain}

\author{Hugo Aramberri}
\affiliation{Institut de Ci\`encia de Materials de Barcelona (ICMAB--CSIC)
             Campus de Bellaterra, 08193 Bellaterra, Barcelona, Spain}

\author{Jorge \'I\~{n}iguez}
\affiliation{Materials Research and Technology Department,
             Luxembourg Institute of Science and Technology,
             41 rue du Brill, L-4422 Belvaux, Luxembourg}

\author{Riccardo Rurali}
\affiliation{Institut de Ci\`encia de Materials de Barcelona (ICMAB--CSIC)
             Campus de Bellaterra, 08193 Bellaterra, Barcelona, Spain}
\email{rrurali@icmab.es}

\date{\today}

\begin{abstract}
We demonstrate a fully electric control of the heat flux, which can
be continuously modulated by an externally applied electric
field in PbTiO$_3$,  a prototypical ferroelectric perovskite,
revealing the mechanisms
by which experimentally accessible fields
can be used to tune the
thermal conductivity by as much as 50\% at room temperature.
\end{abstract}

\maketitle

Our current ability to control heat transport in insulators is rather
limited and mostly consists in modulating the amount of scattering
experienced by the heat carrying phonons~\cite{CahillJAP03}.  This
approach is normally pursued by designing systems with tailor-made
boundaries~\cite{HicksPRB93,LiAPL03}, defect
distributions~\cite{TobererARMS12,CahillAPR14}, or periodical
sequences of different materials or
nanostructuring~\cite{LuckyanovaScience12,
  RavichandranNatMat13,MaldovanNatMat15}, as in superlattices and
phononic crystals. These strategies allow targeting a given thermal
conductivity, which can sometimes result in some degree of thermal
rectification~\cite{TerraneoPRL02,
  ChangScience06,KobayashiAPL09,RuraliPRB14,CartoixaNL15}. Nevertheless,
alternative approaches enabling a {\em dynamical} modulation of the
thermal conductivity are seldom tackled because of the subtleties related
with phonon manipulation.

The intrinsic difficulty in manipulating phonons is often ascribed to
the fact that they do not posses a net charge or a mass; thus, it is
difficult to control their propagation by means of external
fields~\cite{LiRMP12}. However, this is not always the
case~\cite{GroetzingerNature35,BerryJCP49,GairolaJPSJ77,QinNanoscale17}. 
Insulators or semiconductors often feature polar phonons
--which typically involve atoms with different charges, and have a
vibrating electric dipole associated to them-- that can be acted upon
by an external electric field, to {\em harden} or {\em soften} them,
which should result in a modulation of the thermal
conductivity. Further, the structural dielectric response of an
insulator, which is mediated by these very polar modes, may have
significant effects in the entire phonon spectrum, via anharmonic
couplings, and further affect the conductivity. Here we exploit this
simple, yet almost unexplored, idea. We show that the thermal conductivity
can indeed be controlled by an external applied electric field, and
that this effect leads to a genuine thermal counterpart of the field
effect in usual electronic transistors. Indications that such an
electrophononic effect can be obtained experimentally have been
previously reported in SrTiO$_3$ at very low temperatures~\cite{BarrettPRB70,HuberPRB00}

\begin{figure}[t]
\includegraphics[width=1.0\linewidth]{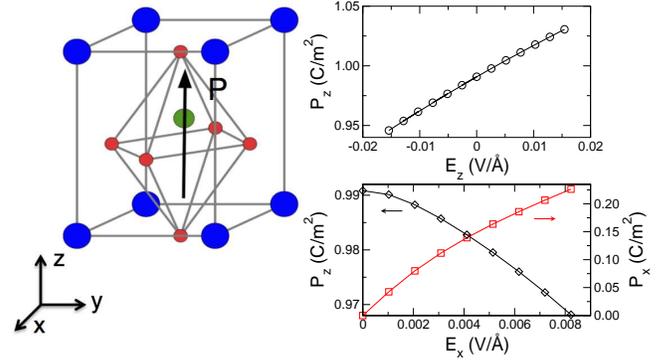}
\caption{(Left) Sketch of a PbTiO$_3$ unit cell. Pb, Ti, and O atoms
  are represented by blue, green, and red spheres, respectively.
  (Right) Polarization as a function of a parallel/antiparallel
  electric field, $E_z$ and of a perpendicular field, $E_x$; in the
  latter case we display both $P_x$ and $P_z$, whose increase/decrease
  allows appreciating the rotation of ${\mathbf P}$.  }
\label{fig:structs}
\end{figure}

We consider PbTiO$_3$ (PTO), a ferroelectric (FE) perovskite that
below a Curie temperature $T_{\rm C} \approx 760$~K has a spontaneous
electric polarization ${\mathbf P}$ associated to the off-centering of
the cations with respect to the surrounding oxygen
atoms~\cite{LinesPR69,rabe-book2007}. PTO's FE phase is tetragonal,
with ${\mathbf P} = P_{z}(0,0,1)$ lying along one of the (pseudo)cubic
directions of the perovskite lattice, as sketched in
Figure~\ref{fig:structs}. By applying electric fields above the
so-called coercive field $E_{\rm coe}$ (which typically lies in the
10$^{6}$--10$^{7}$~V/cm range), it is possible to reverse such a polar
distortion, even in small (nanometric) regions, so that FE domains can
be written. Interestingly, recent works show that juxtaposed domains
with different orientations of ${\mathbf P}$ can be used as phonon
switches~\cite{SeijasBellidoPRB17} and phonon
polarizers~\cite{RoyoPRM17}. Here we consider $E<E_{\rm coe}$,
exploiting the fact that, like most FE materials, PTO displays a
rather large structural (dielectric) response to even moderate applied
fields.

All our simulations of PTO are carried out within
  second-principles density-functional theory (SPDFT) as implemented
  in the {\sc SCALE-UP}
  code~\cite{WojdelJPCM13,GarciaFernandezPRB16}. SPDFT has a
demonstrated predictive power for the key structural, vibrational and
response properties of FE perovskite
oxides~\cite{ZubkoPRL07,WojdePRL14,WojdelPRB14,ZubkoNature16,ShaferPNAS18}. Further, as most
first-principles approaches, SPDFT reproduces accurately the
vibrational and response properties of PTO~\cite{WojdelJPCM13}, which
are closely related to the quantities discussed here; hence, we expect
our results to be quantitatively accurate. For more details on the
used SPDFT methods and the technicalities of our calculations, please
see the Supplemental Material (SM)~\cite{suppl}.
We also observe that, although our simulations are based on a model 
constructed to reproduce the behavior of bulk PbTiO$_3$, previous 
studies showed that the same model works well in a variety of conditions 
that differ from those considered to compute its parameters. This has 
been amply demonstrated, thanks e.g. to the application of the same 
PbTiO$_3$ model employed in this study, in a variety of investigations 
of PbTiO$_3$/SrTiO$_3$ superlattices~\cite{ZubkoNature16,DamodaranNatMar17,ShaferPNAS18}, 
which include successful comparisons with first-principles 
calculations and literature~\cite{Aguado-PuentePRB12} as well as with 
experiment. Hence, the employed models are {\em transferable} to treat 
the most common thin film and superlattice geometries.

\begin{figure}[t]
\includegraphics[width=1.0\linewidth]{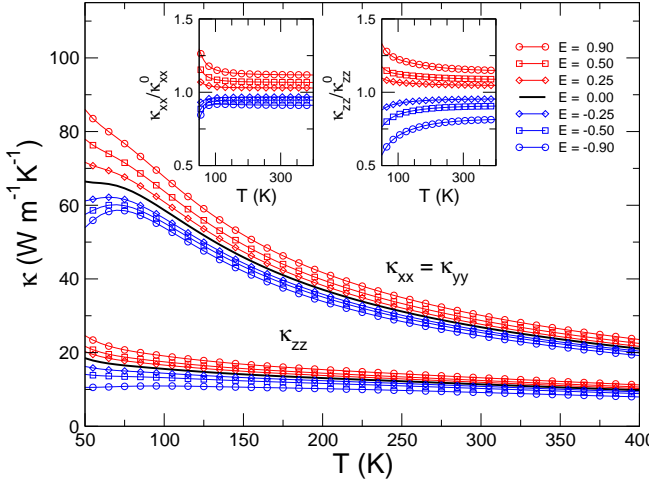}
\caption{Thermal conductivity as a function of temperature for
  different values of the parallel ($E_{z}>0$) and antiparallel
  ($E_{z}<0$) electric field. The inset shows the relative change of
  the thermal conductivity as the ratio of its value with and without
  external field, $\kappa_{xx}/\kappa_{xx}^0$ (left) and
  $\kappa_{zz}/\kappa_{zz}^0$ (right). Fields are given in unit of the
  parallel coercive field, $E_{{\rm coe},z}$}
\label{fig:kappa}
\end{figure}

For the calculation of the thermal conductivity tensor, we proceed as
follows. For each applied field, we first relax the structure by means of a Monte
Carlo simulated annealing, automatically accounting for all dielectric
and piezoelectric effects that may impact the thermal conductivity~\cite{SahooJMC12}. 
Then, we calculate the
second-order interatomic force constants (IFCs) by finite differences
in a $8 \times 8 \times 8$ supercell~\cite{TogoSM15}. We use the same
supercell to compute third-order IFCs~\cite{LiCPC14}, considering
interactions up fourth (twelfth) nearest-neighbors for parallel
(perpendicular) fields, which we check provides good convergence. We
then use the IFCs to calculate the anharmonic scattering rates and
solve numerically the linearized Boltzmann Transport Equation (BTE),
employing the iterative method implemented in the ShengBTE code
\cite{LiCPC14} on a $8 \times 8 \times 8$ ${\bf q}$-point
grid. Scattering from isotopic disorder is accounted for within the
model of Tamura~\cite{TamuraPRB83}.

The lattice thermal conductivity is then obtained as
\begin{equation}
\kappa_{ij} = \sum_{\lambda} \kappa_{ij,\lambda} = C \sum_\lambda
f_\lambda (f_\lambda + 1 ) (h\nu_\lambda)^2 v_{i,\lambda}F_{j,\lambda}
,
\label{eq:kappa}
\end{equation}
where $i$ and $j$ are the spatial directions $x$, $y$, and $z$.
$C^{-1} = k_B T^2 \Omega N$, where $k_B$, $h$, $T$, $\Omega$ and $N$
are, respectively, Boltzmann's constant, Planck's constant, the
temperature, the volume of the 5-atom unit cell, and the number of
${\bf q}$-points. The sum runs over all phonon modes, the index
$\lambda$ including both ${\bf q}$-point and phonon band. $f_\lambda$
is the equilibrium Bose-Einstein distribution function, and
$\nu_\lambda$ and $v_{i,\lambda}$ are, respectively, the frequency and
group velocity of phonon $\lambda$. The mean free displacement
$F_{j,\lambda}$ is initially taken to be equal to $\tau_\lambda
v_{j,\lambda}$, where $\tau_\lambda$ is the lifetime of mode $\lambda$
within the relaxation time approximation (RTA). Starting from this
guess, the solution is then obtained iteratively and $F_{j,\lambda}$
takes the general form $\tau_\lambda
(v_{j,\lambda}+\Delta_{j,\lambda})$, where the correction
$\Delta_\lambda$ captures the changes in the heat current
  associated to the deviations in the phonon populations
  computed at the RTA level~\cite{LiPRB12,TorresPRB17}.

\begin{figure}[t]
\includegraphics[width=1.0\linewidth]{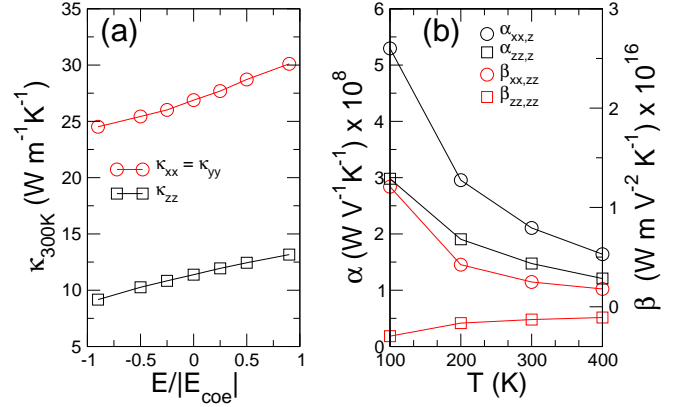}
\caption{(a)~Dependence on the electric field, $E_z$, of the room 
         temperature thermal conductivity. (b)~Coefficients
         ${\bf \alpha}$ and ${\bf \beta}$ of Eq.~\ref{eq:response} 
         as a function of temperature for fields applied parallel 
         to the ${\mathbf P}$ vector.
         }
\label{fig:alpha-beta}
\end{figure}

Our calculations thus yield $\kappa_{ij}$ as a function of applied
field and temperature, and we fit our results to
\begin{equation}
\kappa_{ij}(T,{\mathbf E}) = \kappa_{ij}^{0}(T) + \sum_{k}
\alpha_{ij,k}(T) E_{k} + \sum_{kl} \beta_{ij,kl}(T) E_{k}E_{l} \; ,
\label{eq:response}
\end{equation}
where we introduce the thermal-response tensors $\boldsymbol{\alpha}$
and $\boldsymbol{\beta}$, $\boldsymbol{\kappa}^{0}$ being the
conductivity at zero field. Note that, because of the high tetragonal
symmetry of PTO's FE phase ($P4mm$ space group), the number of
independent tensor components in Eq.~(\ref{eq:response}) is small. For
example, we have $\kappa^{0}_{ij} = \delta_{ij} \kappa^{0}_{ii}$, and
$\kappa^{0}_{xx} = \kappa^{0}_{yy} \neq \kappa^{0}_{zz}$. Here we
focus on the behavior of $\kappa_{xx}$, $\kappa_{yy}$, and
$\kappa_{zz}$ as a function of fields parallel (along $z$) and
perpendicular (along $x$) to $P_{z}$. We thus calculate $\alpha_{xx,z}
= \alpha_{yy,z}$ and $\alpha_{zz,z}$, noting that $\alpha_{ii,x} =
\alpha_{ii,y} = 0$ by symmetry; and we also calculate $\beta_{xx,xx}$,
$\beta_{yy,xx}$, and $\beta_{zz,xx}$, as well as $\beta_{xx,zz} =
\beta_{yy,zz}$ and $\beta_{zz,zz}$.

To explore the linear and non-linear responses, we consider field
values in a range up to 90\% of the theoretical $E_{\rm
  coe}$. Working with an idealized monodomain PTO state with $P_{z} >
0$, our predicted coercive fields are $E_{{\rm coe},z} \approx -1.5
\cdot 10^{8}$~V/m (to reverse $P_{z}$ to $-P_{z}$) and $E_{{\rm
    coe},x} \approx 8.2 \cdot 10^{7}$~V/m (to rotate from $P_{z}$ to
$P_{x}$, a symmetry-equivalent $x$-polarized FE phase). These fields
are relatively large when compared with experimental values, an issue
that is typical of first-principles works on FE switching
\cite{XuNatComm17} and which is probably related, e.g., to
  the absence of nucleation centers for the polarization reversal
(defects, interfaces) in the simulations. This matter is
not important here. Incidentally, note that it is
  customary to apply fields as large as these ones to FE thin films,
  using voltages of a few hundreds meV.

Let us discuss first the response to fields ${\mathbf E} = E_{z}
(0,0,1)$, which can be parallel ($E_{z}>0$) or antiparallel
($E_{z}<0$) to the electric polarization $P_{z} > 0$ (see the SM for
the hysteretic response). Figure~\ref{fig:kappa} shows the thermal
conductivity components $\kappa_{xx}$ and $\kappa_{zz}$, as a function
of temperature, for several values of $E_{z}$. Let us first note that
the zero-field conductivities feature a considerable anisotropy, with,
e.g., $\kappa_{xx}^{0} = 26.9$~W~m$^{-1}$K$^{-1}$ and $\kappa_{zz}^{0}
= 11.4$~W~m$^{-1}$K$^{-1}$ at room temperature ($T_{\rm room}$). This
is a direct consequence of the FE distortion along $z$, and suggests
that, if the electric field is able to affect the polarization
considerably, it will also have a significant effect in the
conductivity. This is indeed what we find. As shown in
Figure~\ref{fig:kappa}, parallel fields yield an increase of both
$\kappa_{xx}$ and $\kappa_{zz}$, while antiparallel fields cause a
decrease. To better appreciate this effect we plot the relative
variation of the thermal conductivities,
$\kappa_{xx}/\kappa_{xx}^0$ and $\kappa_{zz}/\kappa_{zz}^0$ in the
inset.

Further insight can be gathered from
  Figure~\ref{fig:alpha-beta}a, which shows the variation of both
$\boldsymbol{\kappa}$ components as a function of $E_{z}$ at $T_{\rm
  room}$. The obtained smooth behavior can be easily fitted using the
quadratic expression in Eq.~(\ref{eq:response}), and
Figure~\ref{fig:alpha-beta}b shows the $T$-dependence of the
corresponding $\boldsymbol{\alpha}$ and $\boldsymbol{\beta}$
coefficients. The linear response clearly dominates, with
room-temperature values of $\alpha_{zz,z} = 1.47 \cdot
10^{-8}$W~V$^{-1}$K$^{-1}$ and $\alpha_{xx,z}= 2.11 \cdot
10^{-8}$~W~V$^{-1}$K$^{-1}$. As regards the magnitude of the effect,
for $E_{z} = 0.5\times E_{{\rm coe},z}$ at $T_{\rm room}$ we obtain
changes of about 7\% and 9\% in $\kappa_{xx}$ and $\kappa_{zz}$,
respectively. These large effects are ultimately a consequence of
PTO's considerable structural response to the applied fields, as
evidenced by the variation of $P_{z}$ shown in
Figure~\ref{fig:structs}. The corresponding lattice contribution to
the dielectric susceptibility is about 31.

\begin{figure}[t]
\includegraphics[width=1.0\linewidth]{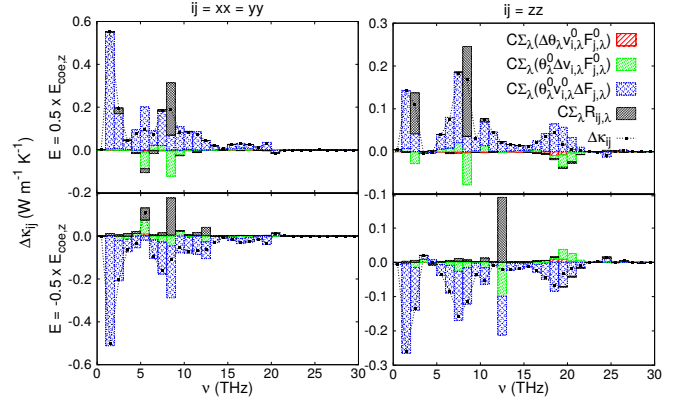}
\caption{The different terms of Eq.~\ref{eq:split} in the case
         of a parallel field, $E_{z} = 0.5 \times E_{{\rm coe},z}$
         (upper row) and of an antiparallel field, $E_{z} = -0.5
         \times E_{{\rm coe},z}$ (bottom row). The dots indicate
         the variation of the total contribution to the thermal
         conductivity in a given frequency interval, i.e. $\Delta
         \kappa_{ij}$ in Eq.~\ref{eq:split}.
         }
\label{fig:split}
\end{figure}

To gain further insight into these results, we find it convenient to
analyze Eq.~(\ref{eq:kappa}) in the following way. First, we group all
the terms that are explicitly dependent on the phonon frequencies by
introducing $\theta_{\lambda} = f_\lambda (f_\lambda + 1 )
(h\nu_\lambda)^2$, and write the field-induced change of $\kappa_{ij}$
as
\begin{equation}
\begin{split}
  \Delta \kappa_{ij} = & \;\; \kappa_{ij} - \kappa^{0}_{ij} =
  \sum_{\lambda} \Delta \kappa_{ij,\lambda}\\ = & \;\; C
  \sum_{\lambda} [\Delta\theta_{\lambda}
    v^{0}_{i,\lambda}F^{0}_{j,\lambda} + \theta^{0}_{\lambda} \Delta
    v_{i,\lambda}F^{0}_{j,\lambda} \\ & \;\; + \theta^{0}_{\lambda}
    v^{0}_{i,\lambda} \Delta F_{j,\lambda} + {\cal R}_{ij,\lambda}] \; ,
\end{split}
\label{eq:split}
\end{equation}
where the superscript ``0'' indicates zero-field quantities, with
$\Delta g = g - g^{0}$ for any magnitude $g$. This expression allows
us to readily identify changes that are dominated by only one of the
factors ($\theta_{\lambda}$, $v_{i,\lambda}$, $F_{j,\lambda}$)
entering the mode conductivity, while ${\cal R}_{ij,\lambda}$ captures
any lingering changes. (In the limit of small applied fields ${\cal
  R}_{ij,\lambda}\rightarrow 0$.) Further, we can group the changes in
the mode conductivities in energy intervals, using the zero-field
frequencies to assign specific modes to specific intervals, and thus
plot Figure~\ref{fig:split} to analyze the $E_{z}$-induced changes in
$\kappa_{xx}$ and $\kappa_{zz}$.

Two important observations can be drawn from this figure. On the one
hand, the change of $\kappa_{xx}$ and $\kappa_{yy}$ does not depend on
a particular group of phonons. Rather, the complete spectrum
contributes to it, in a way that is rather homogeneous. Thus, for
example, we have $\Delta \kappa_{zz} > 0$ for $E_{z} > 0$, where the
total positive change is the result of a majority of phonons having
positive $\Delta \kappa_{zz,\lambda} > 0$ contributions. (Also, note
the approximate symmetry of the results for $+E_{z}$ and $-E_{z}$,
which is consistent with the dominant linear effect.) On the other
hand, for most of the phonon spectrum, it is the change in mean free
displacements that dominates the variation of the conductivity.

We can better understand the changes in $F_{j,\lambda}$ as
follows. First, we can simplify our discussion by noting that
$F_{j,\lambda} = \tau_\lambda (v_{j,\lambda}+\Delta_{j,\lambda})
\approx \tau_\lambda v_{j,\lambda}$, as we observe that the correction
to the RTA is small, typically below a 10\%. Then, we find that the
changes in phonon lifetimes dominate over the variations of the group
velocities, which is consistent with the relatively modest impact of
the $\Delta v_{j,\lambda}$ term shown in
Figure~\ref{fig:split}. Further, as described in the SM and
Ref.~\onlinecite{LiCPC14}, we have $\tau_{\lambda}^{-1} \sim
f_{\lambda'}\times (\nu_{\lambda}\nu_{\lambda'}\nu_{\lambda''})^{-1}$,
where $\lambda'$ and $\lambda''$ label modes that interact with
$\lambda$ via a three-phonon scattering process. Hence, for example,
if most phonons were to harden under application of a field $E_{z} >
0$, the phonon frequencies $\{\nu_{\lambda}\}$ would generally
increase and the populations $\{f_{\lambda}\}$ decrease, which would
yield an increase of the lifetimes $\{\tau_{\lambda}\}$. This is
precisely what we have in our calculations, as the average phonon
frequency changes from $\bar{\nu}^{0} = 9.96$~THz to $\bar{\nu} =
10.05$~THz for $E_{z} = 0.5 \times E_{{\rm coe},z}$, resulting in
generally longer lifetimes and larger thermal conductivity. In
contrast, for $E_{z} = - 0.5 \times E_{{\rm coe},z}$ we obtain
$\bar{\nu} = 9.87$~THz, with generally shorter lifetimes and greater
thermal resistance~\cite{freq}. Indeed, we find that this is the
dominant effect explaining our results for $\kappa_{xx}$ and
$\kappa_{zz}$ under fields that are (anti)parallel to the polarization
$P_{z}$.

The fact that most of PTO's phonon bands become harder for $E_{z} > 0$
(softer for $E_{z} < 0$) may seem surprising at first; yet, we believe
it can be rationalized as follows. According to our simulations, the
application of a parallel field has two main effects. On one hand, the
cell volume grows moderately. For example, we get $\Omega/\Omega^{0} =
1.0018$ for $E_{z} = 0.5 \times E_{{\rm coe},z}$, which is a
consequence of a dominant piezoelectric effect. The increased volume
alone should result in a general softening (reduction) of the phonon
frequencies, which is the usual behavior corresponding to a positive
Gr\"uneinsen parameter. On the other hand, $P_{z}$ grows for $E_{z} >
0$, and the stronger polar distortion can also be expected to have an
impact on the phonon frequencies. More precisely, in the field of
phase transitions in perovskites, it is generally observed that
different distortions of the cubic perovskite structure tend to {\em
  compete} with each other, implying that the condensation of one
(e.g., the polar distortion) tends to {\em harden} the others, thus
increasing the associated phonon frequencies. (See
Ref.~\onlinecite{WojdelJPCM13,MercyNatComm17}) Our results suggest
that this effect is dominant in PTO.

Since we attribute the changes in conductivity under $E_{z}$-field to
a general hardening/softening of the phonon spectrum, it may seem
strange to note in Figure~\ref{fig:split} that the changes associated
to the $\Delta\theta_{\lambda}$ term [Eq.~(\ref{eq:split})]
are negligible (in fact, they are barely visible in the
  figure). Yet, note that, in this term, the variations of
frequencies and populations tend to cancel each other, yielding a
relatively small net effect.

\begin{figure}
\includegraphics[width=1.0\linewidth]{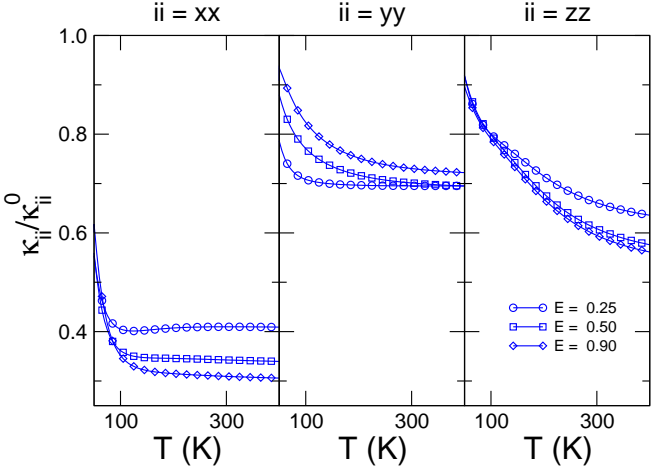}
\caption{Ratio of the thermal conductivity with and without a
  perpendicular external field, $\kappa_{xx}/\kappa_{xx}^0$ (left),
  $\kappa_{yy}/\kappa_{yy}^0$ (center), and
  $\kappa_{zz}/\kappa_{zz}^0$ (right), as a function of temperature.
}
\label{fig:kappa_rel_perp}
\end{figure}

Let us now move to the case in which we apply a field ${\mathbf E} =
E_{x}(1,0,0)$, perpendicular to the polarization,
  $P_{z}$. We consider $E_{x} > 0$, noting that this situation is
equivalent by symmetry to the application of $E_{x} < 0$ or fields
along $y$. Figure~\ref{fig:kappa_rel_perp} summarizes our results,
which feature a very large decrease of all the tensor
components. Thus, for example, for $E_{x} = 0.25 \times E_{{\rm
    coe},x}$ at $T_{\rm room}$, we get $\kappa_{xx}/\kappa_{xx}^{0} =
0.41$, $\kappa_{yy}/\kappa_{yy}^{0} = 0.70$, and
$\kappa_{zz}/\kappa_{zz}^{0} = 0.66$. This dramatic enhancement of the
thermal resistance translates into very large values of the quadratic
response ${\mathbf \beta}$, as we obtain $\beta_{xx,xx} = -1.39 \cdot
10^{-12}$~WmV$^{-2}$K$^{-1}$, $\beta_{yy,xx} = -2.26 \cdot
10^{-13}$~WmV$^{-2}$K$^{-1}$, and $\beta_{zz,xx} = -6.51 \cdot
10^{-14}$~WmV$^{-2}$K$^{-1}$ at $T_{\rm room}$.

\begin{figure}[t]
\includegraphics[width=1.0\linewidth]{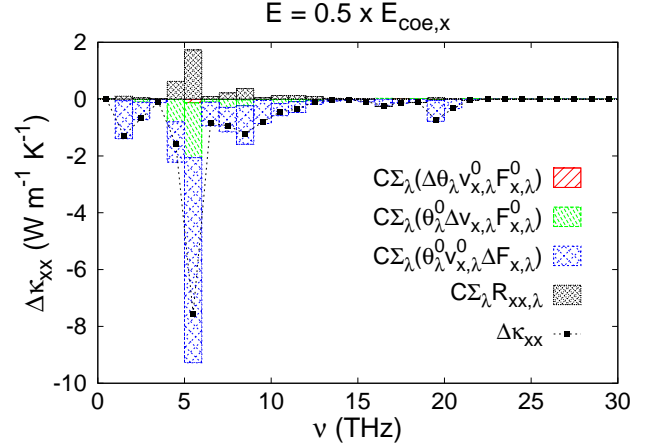}
\caption{The different terms of Eq.~\ref{eq:split} in the case
         of a perpendicular field, $E_{x} = 0.5 \times E_{{\rm coe},x}$
         The dots indicate the variation of the total contribution to
         the thermal conductivity in a given frequency interval,
         i.e. $\Delta \kappa_{xx}$ in Eq.~\ref{eq:split}.
         }
\label{fig:split_perp}
\end{figure}

Figure~\ref{fig:split_perp} shows the analysis based on
Eq.~(\ref{eq:split}), applied to the change in $\kappa_{xx}$ at
$T_{\rm room}$ for a field $E_{x} = 0.5 \times E_{{\rm coe},x}$, which
is a representative case. As above, we find that the total $\Delta
\kappa_{xx}$ is the result of contributions spanning the whole phonon
spectrum, and dominated by the changes in mean free paths. Also as
above, we find that it is the change in the phonon lifetimes what
controls $\Delta F_{j,\lambda}$; yet, at variance with the case of the
$E_{z}$-fields, the present effect cannot be attributed to a general
shift of frequencies. Indeed, we find that the $E_{x}$-field tends to
harden the phonon spectrum (e.g., we obtain $\bar{\nu} = 10.02$~THz
for $E_{x} = 0.5 \times E_{{\rm coe},x}$). According to our above
argument to explain the response to $E_{z}$-fields, the larger
frequencies should result in longer lifetimes and an increased
conductivity; yet, the effect of the perpendicular fields
  is just the opposite, with increased resistivity. Interestingly, a
further analysis of our results reveals that, in this case, the
field-dependence of the lifetimes is dominated by the three-phonon
scattering matrix $V_{\lambda\lambda'\lambda''}$, which controls the
phonon decay as $\tau_{\lambda}^{-1} \sim
|V_{\lambda\lambda'\lambda''}|^{2}$~\cite{LiCPC14}. More specifically,
we find that the $E_{x}$-field activates a large number of new scattering
processes due to the symmetry breaking that it causes. (An
  $E_{z}$-field does not change the symmetry of PTO's
$P_{z}$-polarized phase, and the proliferation of scattering events
does not occur in that case.)  This effect affects the whole phonon
spectrum, and its magnitude naturally scales with the structural
symmetry breaking caused by $E_{x}$, which is rather considerable
given the large dielectric response of PTO to such a perturbation (for
the corresponding susceptibility we obtain $\chi_{xx} \approx 304$;
see Figure~\ref{fig:structs}). Such a strong response to a transversal
field is related to the ``easy polarization rotation'' that is
well-known in FE perovskite oxides \cite{FuNature00,BellaichePRL00}.

Interestingly, we also observe a field-induced coupling of the $x$ and
$z$ directions -- i.e., those along which the initial ${\mathbf P}$
and the applied $E_{x}$-field are oriented -- in the thermal
conductivity tensor. We obtain values of $\kappa_{xz}$ and
$\kappa_{zx}$ that are not negligible, of the same order of those in
porous~\cite{GeseleJPD97,DettoriPRB15} or amorphous materials~\cite{ZinkPRL06}.
Their temperature dependence for applied fields is shown in the SM.
This non-zero components imply that, e.g., a thermal gradient along
$x$ results in a heat flux, not only along $x$, but also along $z$.

In conclusion, we have reported evidence of the coupling between
electric field and thermal conductivity in a ferroelectric perovskite.
We have shown that an electric field perpendicular to the spontaneous
polarization greatly increases the thermal resistivity, the underlying
physical mechanism being the breaking of the symmetry of the lattice,
which activates new scattering processes with a concomitant reduction
of the lifetimes of phonons throughout the vibrational spectrum.  On
the other hand, for parallel fields that do not activate new
scattering processes, we observe a linear variation of the thermal
conductivity, which can grow or decrease depending on the sign of the
applied field. This linear effect is controlled by the overall
hardening/softening of the phonon modes. The predicted behaviors open
the way to a fully-electric control of phonon transport.
As the underlying
physical principle is the manipulation of polar modes, these results
can potentially be extended to a broader class of materials, possibly
with even larger responses. Finally, we note that the
symmetry breaking that leads to the largest changes in the thermal
conductivity can also be achieved in other ways, 
such as mechanical strains, that do not involve electric fields.
Implementations of these concepts in a realistic device will have 
to take into account that substrates and additional layers will 
provide alternative heat transport channels and that the thermal 
contact resistance~\cite{YazjiNR15,YazjiSST16} may complicate our 
taking advantage of the controllable transport properties of the 
ferroelectric layer. These issues, however, admittedly fall beyond 
the scope of the present manuscript.

\begin{acknowledgments}
J.A.S.-B., H.A. and R.R. acknowledge financial support by the Ministerio
de Econom\'ia, Industria y Competitividad (MINECO) under Grants No.
FEDER-MAT2013-40581-P, No. FEDER-MAT2017-90024-P and the Severo Ochoa
Centres of Excellence Program under Grant No. SEV-2015-0496 and by the
Generalitat de Catalunya under Grants No. 2014 SGR
301 and No. 2017 SGR 1506. J.\'I. acknowledges the support of the Luxembourg
  National Research Fund through the PEARL program (Grant No. FNR/P12/4853155 COFERMAT).
\end{acknowledgments}




%

\end{document}